# Towards Reproducible Research: Automatic Classification of Empirical Requirements Engineering Papers

Clinton Woodson, Jane Huffman Hayes, Sarah Griffioen
Department of Computer Science
University of Kentucky
Lexington, Kentucky, United States of America
clint.woodson@uky.edu, hayes@cs.uky.edu, sarah.griff@uky.edu

## ABSTRACT

Research must be reproducible in order to make an impact on science and to contribute to the body of knowledge in our field. Yet studies have shown that 70% of research from academic labs cannot be reproduced. In software engineering, and more specifically requirements engineering (RE), reproducible research is rare, with datasets not always available or methods not fully described. This lack of reproducible research hinders progress, with researchers having to replicate an experiment from scratch. A researcher starting out in RE has to sift through conference papers, finding ones that are empirical, then must look through the data available from the empirical paper (if any) to make a preliminary determination if the paper can be reproduced. This paper addresses two parts of that problem, identifying RE papers and identifying empirical papers within the RE papers. Recent RE and empirical conference papers were used to learn features and to build an automatic classifier to identify RE and empirical papers. We introduce the Empirical Requirements Research Classifier (ERRC) method, which uses natural language processing and machine learning to perform supervised classification of conference papers. We compare our method to a baseline keyword-based approach. To evaluate our approach, we examine sets of papers from the IEEE Requirements Engineering conference and the IEEE International Symposium on Software Testing and Analysis. We found that the ERRC method performed better than the baseline method in all but a few cases.

## 1 INTRODUCTION

Innovative research is a vital part of moving the requirements engineering industry forward, spurring the development of novel, faster, and better techniques. While current emphasis is placed on "greenfield" research, there is a decline in reproducible research, regardless of whether the research being reproduced is greenfield or not. According to Popper [2], "Non-reproducible single occurrences are of no significance to science." Studies show that up to 70% of academic research isn't able to be reproduced, which "represents a tremendous amount of wasted effort and money [1]." If research cannot be reproduced, there isn't an efficient way to determine its validity, which results in it going unused.

Recent work funded by the National Science Foundation developed a research framework called TraceLab [3]. TraceLab is designed to "provide an experimental environment in which researchers can design and execute experiments [3]." While TraceLab allows researchers to easily reproduce experiments, it should first be determined if the work in a given research paper even has the possibility of being reproduced. While the ultimate goal of our research is to be able to quickly determine whether an experiment or study in a paper can be reproduced, this paper addresses antecedent questions to support that objective.

The first step in our overall process is to determine whether a paper is related to Requirements Engineering (RE), as we are focused on replicating requirements engineering research. Next, we need to determine whether the RE paper is empirical. We define an empirical paper as one that is based on observed and measured phenomena, where results are analyzed and conclusions are drawn.

We start by applying natural language processing (NLP) techniques. Once the text of the research papers has been extracted and processed, we apply two methods for determining features of the papers. The first is a baseline method which uses the frequency of certain key words. The other method, the Empirical Requirements Research Classifier (ERRC), uses supervised learning as the basis for the features. Both methods have been implemented as TraceLab components.

We built a training set by manually labelling papers from several years' worth of papers from two different conferences (one RE, one not). We then applied popular classification techniques to each model: Weka's [6] implementation of Naïve Bayes [7], J48 [8], and ZeroR [9]. We used precision, recall, and f-measure, as well as the prediction accuracy, to evaluate the ERRC and baseline methods.

The paper is organized as follows. Section II discusses the research method. Section III addresses the study approach, including the threats to validity. The results and analysis are presented in Section IV. Related work is discussed in Section V. Section VI presents our conclusions and future work.

## 2 METHOD

Figure 1 presents a high level overview of our ultimate goal: to automatically identify reproducible empirical requirements engineering papers. The shaded blocks are in the scope of this paper. Toward our goal we developed a method to identify empirical requirements engineering papers, the ERRC.

As can be seen in the figure, we first created a directory for each year of each conference and saved the file for each paper. To support model building, we manually labelled each paper as being empirical RE, non-empirical RE, empirical non-RE, or non-empirical non-RE.

Next, each paper was parsed. The newline and return characters were then removed. This allowed, for example, any phrases spanning multiple lines to be read as a whole. Pre-processing was performed: we replaced all punctuation except apostrophes and dashes with spaces to allow for easier text recognition. We then used a filter to remove stop words. Stop words are common words that don't add meaning to a sentence (i.e., "the," "an," "and"). Finally, numbers were removed from the text.

Once the text was processed, we proceeded with data collection. Each remaining word was shortened to its stem (i.e., "required," "requirements," "requiring" all were stemmed to "requir") and added to the list of stems (unless the stem had already been found, in which case we increased the count for that stem). Stemming words to their morphological root increases the likelihood of similar or related words being matched.

Once data collection was complete, we used two different feature selection methods to build the classification models.

### 2.1 Baseline Method

The baseline approach used a simple term frequency count for developing features of the model. We identified key words from previous RE and empirical papers. Specifically, we identified the top five most frequently used keywords from the most recent RE and ISSTA conferences, which spanned over 200 papers. The term counts of the selected key words were then used to build the baseline model.

### 2.2 ERRC Method

The ERRC method represents a more general approach to paper classification. Unlike the baseline method, which uses only the provided key terms, the ERRC method recorded the frequency of all stemmed terms found in a paper. Once a complete list had been created, the terms were sorted from most frequent to least. The top ten most frequent stemmed terms were then recorded as the unique features for that paper. The result was the ERRC model that can be passed to a classification technique.

### 2.3 Analysis of Methods

To measure the effectiveness of the baseline and ERRC methods, we used Weka to classify the resulting models. Each classification technique, ZeroR, Naïve Bayes, and J48, was applied to each method's model. The models were evaluated using cross validation at 10, 20, 30, and 40 folds. In cross validation, the dataset is divided into *k* subsets. One of the *k* subsets is used for the testing set, while the other *k-1* subsets are used for the training sets. The advantage of this method is that every element gets to be in the testing set exactly once, and in the training set *k-1* times [15]. Each cross validation was run 10 times with a different seed, to randomize the folds. We used the average of the results of all runs to perform analysis.

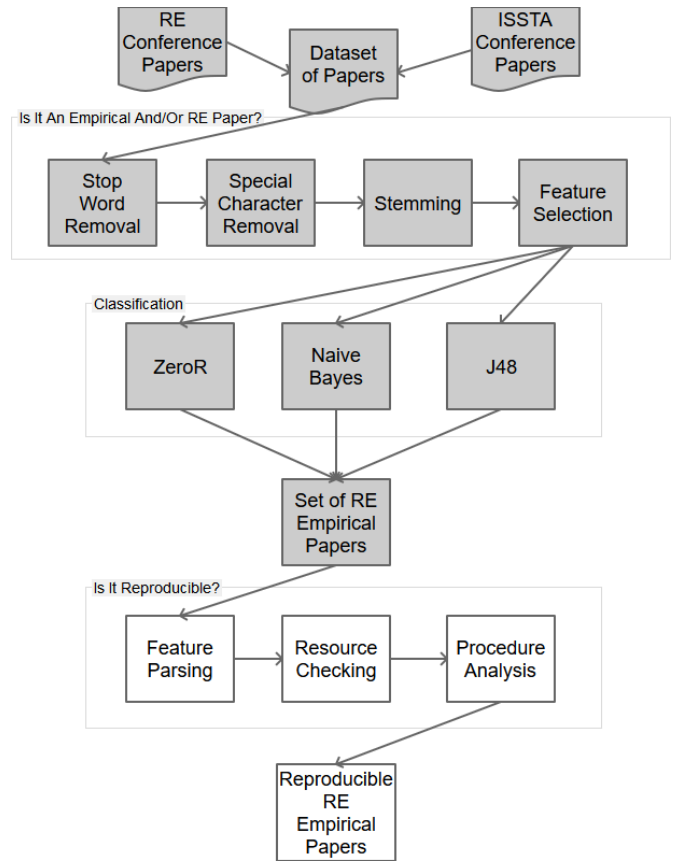

Figure 1: High level overview of approach to identification of reproducible empirical requirements engineering papers.

## 3 EMPIRICAL STUDY

We performed a case study aimed at evaluating whether RE empirical research papers can be automatically classified. We studied two research questions.

**RQ1**: Can NLP features be used to characterize empirical and/or RE papers?

**RQ2**: Can the ERRC method predict paper classifications better than the baseline method?

Studying RQ1 and RQ2 will help determine whether or not a model can be built to classify RE empirical papers. With this question answered, we will be able to move onto our longer term research project of determining if it is possible to classify papers as reproducible.

For this study, we have a null hypothesis

$$H_0 : A_{ERRC} = A_B$$

and an alternate hypothesis

$$H_1 : A_{ERRC} > A_B$$

where A is the accuracy of the method, ERRC is the ERRC method, and B is the baseline method.

### 3.1 Objects of Analysis

For the objects of analysis, we chose conference papers from the IEEE Requirements Engineering (RE) conference and the IEEE International Symposium on Software Testing and Analysis (ISSTA). The RE conference ensures that papers on requirements engineering research are represented; the ISSTA conference ensures that non requirements engineering research is represented. Further, we chose these two conferences since they have both RE and empirical papers.

The breakdown of papers used is shown in Table 1. We chose the years 2000, 2005, and 2015 for RE to represent an even division of years across the past conference offerings. We chose 2000, 2004, and 2015 for ISSTA for the same reason. The papers ranged from 5-10 pages in length, with most of them being 10 pages.

Once a suitable subset of papers was gathered, we manually classified all the papers. To accomplish this, we had one of the co-authors read through the papers, labelling them as RE or not, and empirical or not. Note that there is a good balance of empirical/non empirical and RE/ non RE papers, as can be seen in the bottom row of the table.

**Table1: Conference Papers**

| Year | Empirical | Non-Empirical | RE | Non-RE | Total |
|---|---|---|---|---|---|
| IEEE RE | | | | | |
| 2000 | 7 | 6 | 12 | 1 | 13 |
| 2005 | 21 | 23 | 41 | 3 | 44 |
| 2015 | 29 | 18 | 43 | 4 | 47 |
| | | | | | |
| IEEE ISSTA | | | | | |
| 2000 | 12 | 9 | 1 | 20 | 21 |
| 2004 | 11 | 17 | 1 | 27 | 28 |
| 2015 | 21 | 21 | 0 | 42 | 42 |
| Total | 101 (52%) | 94 (48%) | 98 (50%) | 97 (50%) | 195 (100%) |

### 3.2 Variables and Measures

This section describes the independent and dependent variables of our study.

#### 3.2.1 Independent Variables.
We had two independent variables. First, we varied the feature selection methods, applying a baseline approach and the ERRC. The baseline method was developed to be a simplistic approach to be used as a control method against which to judge the ERRC method.

Second, we used three classification techniques in the study: ZeroR, Naïve Bayes, and J48. We implemented these methods using the Weka Data Mining Software [6].

ZeroR is one of the simplest classification methods. It ignores any predictors, only relying on the target of the data. With its lack of ability to predict anything other than the majority class, it is unhelpful for practical prediction, but is useful for creating a baseline result against which to compare the other techniques.

The Naïve Bayes classifier is built upon the Bayes' theorem. Naïve Bayes uses independent assumptions for the features to predict the classification.

J48 is an open-source Java implementation of the C4.5 algorithm. It builds decision trees from the training set.

#### 3.2.2 Dependent Variables.
We chose accuracy, recall, precision, and f-measure as the dependent variables. Accuracy measures the percent of correctly classified papers. Precision measures how many of the retrieved elements are relevant (how many of the papers that ZeroR indicates are RE papers truly are?). Recall, on the other hand, measures the percentage of true instances that are retrieved (did ZeroR retrieve all the RE papers?). F-measure is the harmonic mean of recall and precision and provides a single measure to represent both.

### 3.3 Study Operation

To perform the feature setup and collection, we used the Apache Lucene and Solr libraries [12]. These Java-based libraries provide text indexing, searching, and advanced analysis/tokenization capabilities. We used these libraries to remove stop words, remove special characters and numbers, and to stem words and count their frequencies.

To help speed up the development and evaluation process, and to allow others to easily reproduce our experiment, we implemented the study as a collection of TraceLab components.

### 3.4 Threats to Validity

The primary threat to external validity in this experiment involved the datasets. Other datasets may be larger, or have different term frequencies. A larger dataset may generate a more diverse classification model. Also, due to the inability and impracticality of building a model that uses every paper from every conference from every field, we mitigated the threat by using papers from several years' worth of two different con-ferences. We cannot claim that our results will generalize to other datasets.

For internal validity, the primary threat involved the manual classification of the training set. To reduce this threat, we had one co-author perform the labelling. These labels were then later corroborated by another co-author. Both co-authors worked independently. The co-authors discussed conflicting labels until agreement was reached. To perform analysis, we used popular and established tools (i.e., Weka).

For construct validity, the primary concern was the dependent variables used to answer the research questions. To address this threat, we used the standard and well accepted measures of accuracy, recall, precision, and f-measure. To minimize conclusion validity threats, we performed statistical analysis to interpret the results.

## 4 RESULTS AND ANALYSIS

All of the tables for our results can be found at www.cs.uky.edu/~hayes.

### 4.1 RE Paper Classification

Figure 2 shows the percent of correctly classified empirical papers using the Naïve Bayes technique. As can be seen, the ERRC method has about a 2% accuracy benefit over the baseline method at 10 folds. Figure 3 shows the same using the J48 technique. The ERRC method outperforms the baseline method by a maximum of 10%. Figures 4 and 5 show the precision for the empirical classification using Naïve Bayes and J48. These show that the ERRC method has increased precision over the baseline method using J48, but not when using Naïve Bayes. Figure 6 shows the recall using J48 for the empirical classification. As can be seen, the ERRC method has a steady value across all numbers of folds, whereas the baseline method varies greatly, with its lowest value under 0.2. Figure 7 shows the f-measure using Naïve Bayes while classifying empirical papers, with the ERRC having almost double the value of the baseline method.

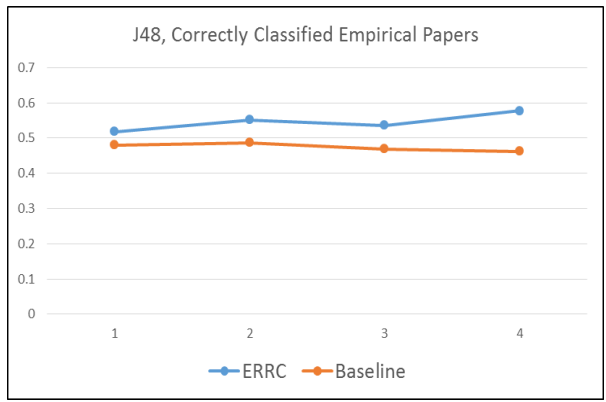

**Figure 3: Percent of Correctly Classifed Empirical Papers using J48. X-Axis = number of Cross validation folds times 10.**

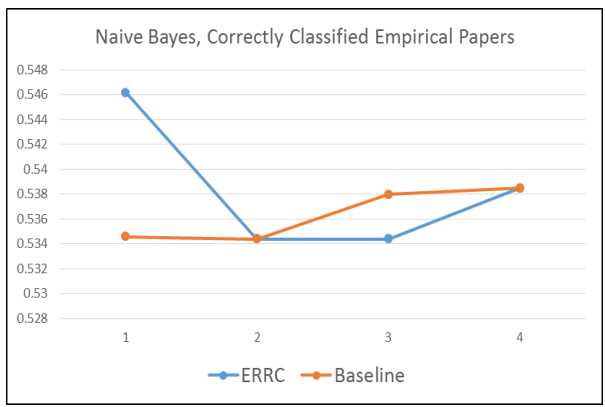

**Figure 2: Percent of Correctly Classifed Empirical Papers using Naïve Bayes. X-Axis = number of Cross validation folds times 10.**

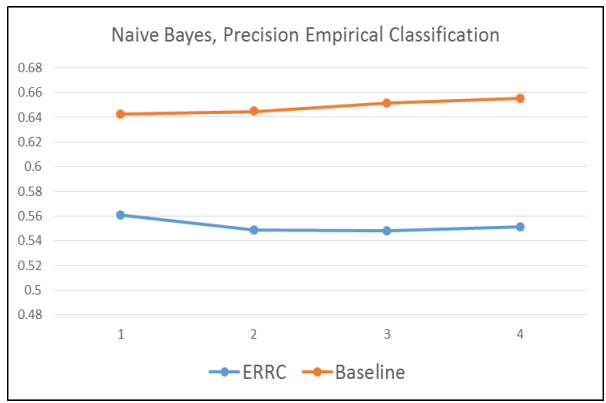

**Figure 4: Naïve Bayes Precision for Classifying Empirical Papers. X-Axis = number of Cross validation folds times 10.**

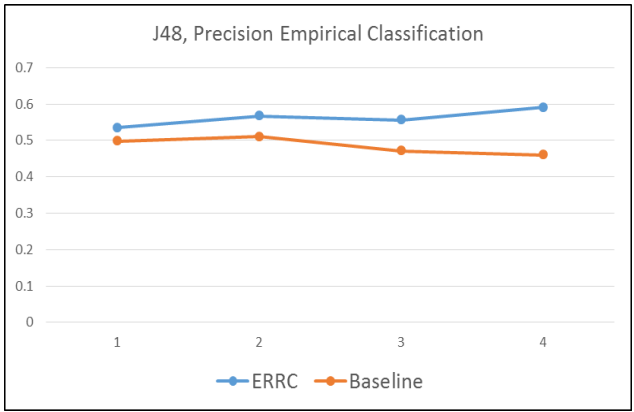

Figure 5: J48 Precision for Classifying Empirical Papers. X-Axis = number of Cross validation folds times 10.

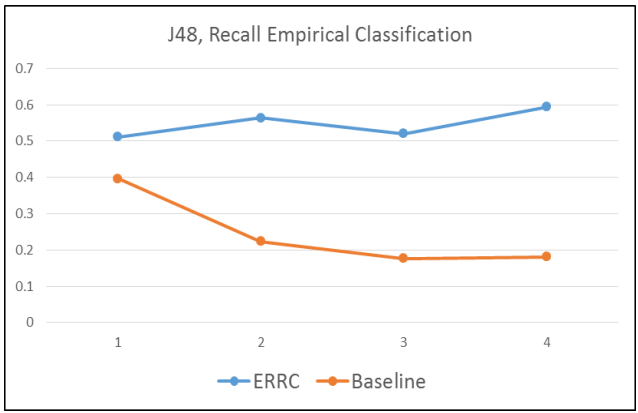

Figure 6: J48 Recall for Classifying Empirical Papers. X-Axis = number of Cross validation folds times 10.

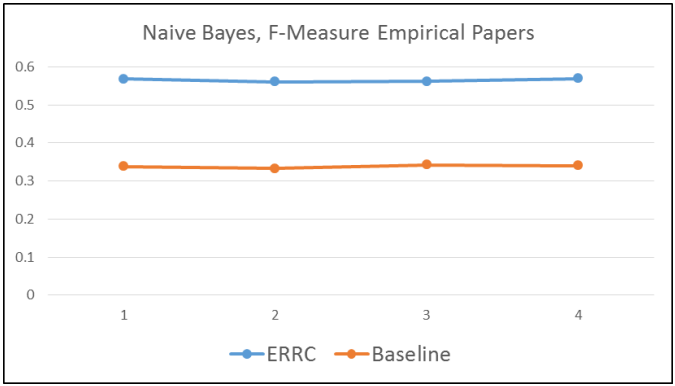

Figure 7: Naïve Bayes f-measure for Classifying Empirical Papers. X-Axis = number of Cross validation folds times 10.

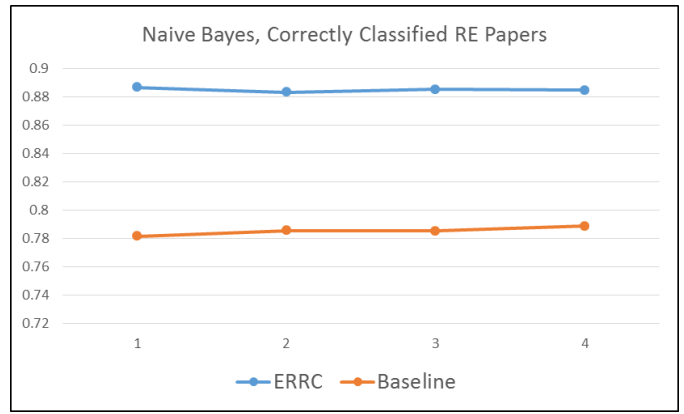

Figure 8: Percent of Correctly Classifed RE Papers using Naïve Bayes. X-Axis = number of Cross validation folds times 10.

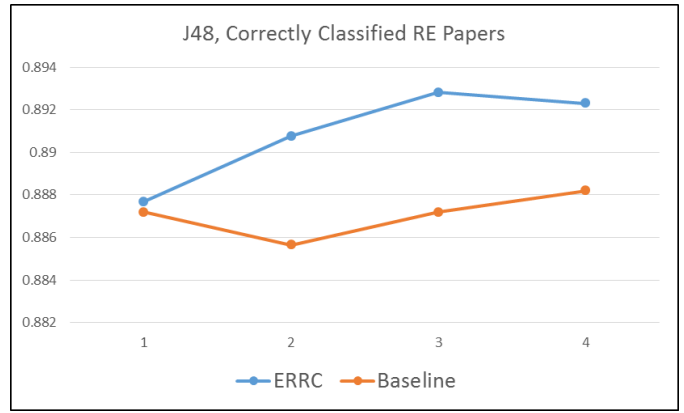

Figure 9: Percent of Correctly Classifed RE Papers using J48. X-Axis = number of Cross validation folds times 10.

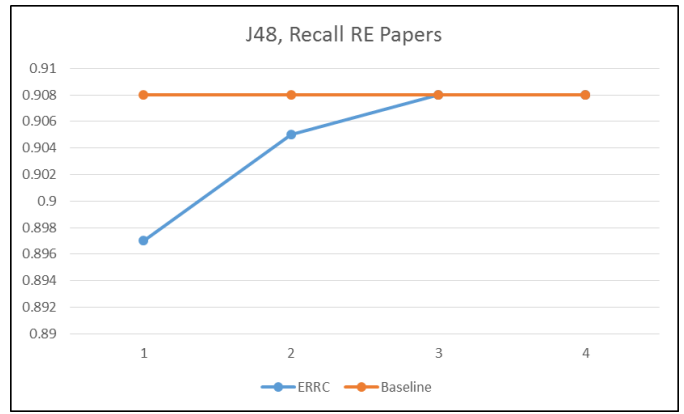

Figure 10: J48 Recall for classifying RE papers. X-Axis = number of Cross validation folds times 10.

*4.2 Empirical Paper Classification*

The percent of correctly classified papers using Naïve Bayes can be seen in Figure 8. As shown, the ERRC method outperforms the baseline method by about 10%. Figure 9 also shows the ERRC method outperforming the baseline method using J48. While the ERRC and baseline methods have close recall, as seen in Figure 10, the ERRC has f-measure roughly 12% higher, which is shown in Figure 11.

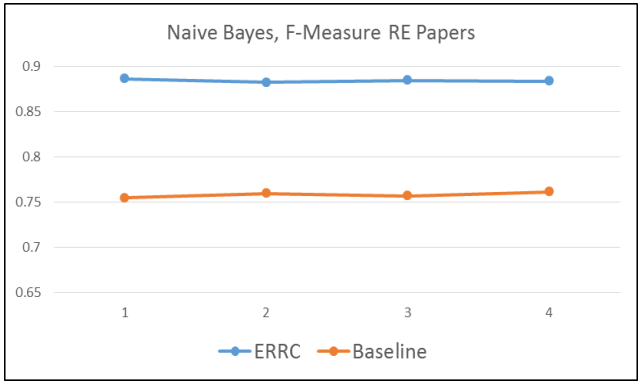

**Figure 11: Naïve Bayes f-measure for classifying RE papers. X-Axis = number of Cross validation folds times 10.**

### 4.3 Analysis

While the ERRC and baseline methods may have similar accuracy classifying empirical papers, the ERRC has about 10% increased performance for classifying RE papers.

Surprisingly, the baseline method outperformed the ERRC method for classifying some empirical papers using Naïve Bayes at higher folds. This result does not mean that the ERRC method is not useful. The ERRC method still has an advantage of automatic modelling over manually assigning terms.

For RE classification, the ERRC method clearly outperforms the baseline method, with a much higher classification accuracy, recall, and f-measure. The baseline method does have a slightly higher precision, though. This higher precision could be due to the baseline method naively predicting an RE classification more often than the ERRC, which would explain the low recall and f-measure values. This is not certain though as there was not an overwhelming majority class in the dataset. Recall from Table I that the training set was well balanced, with 98 RE papers and 97 non-RE papers.

Table 2 shows the one tail t-test statistical analysis of the accuracy, recall, precision, and f-measure from the study. For this study α = 0.05, meaning that there is a 5% or less probability that the results are due to chance. The values in Table 2 which are less than α, and therefore significant, are bolded. As shown in the table, all values are significant except for the Naïve Bayes for empirical papers and J48 for RE papers. To answer RQ1, we can use NLP features to characterize empirical and/or RE papers. Our results show that we can characterize empirical papers with roughly 55% accuracy and RE papers with roughly 89% accuracy. Answering RQ2, the ERRC does predict paper classifications as well as or better than the baseline method. Due to the significance of the results, we can reject our null hypothesis $H_0$ in favor of our alternate hypothesis $H_1$.

**Table 2: Statistical analysis**

|  | Empirical | Requirements |
|---|---|---|
| Accuracy P(T<=t) one tail | | |
| Naïve Bayes | 0.294063822 | **9.22234E-06** |
| J48 | **0.01042203** | **0.022431733** |
| Recall P(T<=t) one tail | | |
| Naïve Bayes | **9.71788E-07** | **4.78814E-06** |
| J48 | **0.009172371** | 0.135325556 |
| Precision P(T<=t) one tail | | |
| Naïve Bayes | **0.000168851** | **0.00188626** |
| J48 | **0.016667758** | **0.000890836** |
| F-Measure P(T<=t) one tail | | |
| Naïve Bayes | **1.46014E-06** | **5.56186E-06** |
| J48 | **0.010584164** | 0.057447213 |

## 5 RELATED WORK

Hayes, Li, and Rahimi [16] discuss the potential that can be achieved in requirements engineering research when the Weka machine learning software suite and the TraceLab project are combined. Towards this goal, they implement a proof of concept in the form of a TraceLab component which uses the Weka classification trees. They demonstrate the usability of their component on two different requirements engineering problems. They also offer insights on using their Tracelab Weka component. Their work relates to this paper as we also use TraceLab and Weka to support our study.

The first defense against software bugs is to develop testable requirements. This allows developers to test that their implementation of a requirement is correct. Hayes et al. [17] examined two datasets with requirement and code artifacts to address testability from the perspective of requirement understandability and quality. Their work relates to this paper in that both use machine learning to automatically classify a textual dataset. We classify research papers, Hayes et al. [17] classify whether a requirement is testable.

Dit, Moritz, Linares-Vasquez, Poshyvanyk, and Cleland-Huang [4] attempt to remedy the problem of software maintenance research studies having difficult to reproduce experiments. They found that studies are hard to reproduce due to a lack of datasets, tools, implementation details, and other varied reasons. This hurdle hinders progress in the field by requiring researchers to devote a significant amount of time to recreating test processes to determine if new techniques truly are an improvement over existing ones. Their research is applicable to our work by attempting to alleviate the difficulty of sharing experiment sources. With the approach of Dit, Moritz, Linares-Vasquez, Poshyvanyk, and Cleland-Huang, components can be developed that can be used to reproduce an experiment on any machine, with little or no setup required on the tester's side.

Millions of apps can be found in the different app stores, and with them billions of reviews for the apps. This large amount of data is a significant source of user feedback that can be used to develop higher quality apps. There is a challenge, though, with sifting through which reviews are relevant or not. Maalej and Nabil [5] discuss several techniques for classifying these reviews into different types. This classification of unstructured text is similar to our research of classifying conference papers.

McCallum and Nigam [7] discuss two different first-order probabilistic model approaches to text classification using the Naïve Bayes assumption: a multi-variate Bernoulli model and a multinomial model. Their [7] results find that the multi-variate Bernoulli method performs better with small vocabulary sizes, but the multinomial method performs better with larger vocabularies. Their work relates to our paper as they also use a Naïve Bayes method to classify the text in their experiment.

## 6 CONCLUSION AND FUTURE WORK

With the quantity of academic research, and concomitantly the number of publications, on the rise, the amount of research that cannot be reproduced has also risen. To be able to determine the reproducibility of an academic research paper, we have worked on determining if a paper is an RE paper, and then whether that paper is an empirical paper. To approach this, we took papers from the IEEE Requirements Engineering and the IEEE International Symposium on Software Testing and Analysis conferences and collected data to build training sets. We built a baseline keyword-based method and our ERRC method to model the academic research, then applied various classification techniques. Our results show that our ERRC method performed approximately 3% better than the baseline method at classifying empirical papers and 12% better when classifying RE papers.

While the ERRC method shows promise, there is definitely room to improve. The first possible improvement would be to expand the stop word list to help further filter out words that add no meaning to the classification of the paper. Along the same lines, other filtering could help narrow the scope of what the paper being classified is about. Possibilities include comparing the paper's text against a dictionary to remove acronyms, project names, and other special words. The potential downside of these approaches could be filtering too much out, thus possibly removing important words. Another way to possibly filter the text of the paper under analysis would be to weight the words based on the location in which they were found in the paper. Words found in an abstract or conclusion could be given more weight than words found in the body of the paper, for example. The reasoning for this is that we hypothesize that words found in those locations would more directly address the content of the paper being analyzed.

Source code and datasets for the study can be found at www.cs.uky.edu/~hayes.

## ACKNOWLEDGMENT

This work was supported in part by NSF grant CCF-1511117.

## REFERENCES

[1] Hahnel, Mark. "Reproducility of Research – A New Standard." Figshare. 14 Aug. 2012. Web. 1 June 2016

[2] Popper, K.R., "Non-reproducible single occurrences are of no significance to science." 1959. The logic of scientific discovery. Hutchinson, London, United Kingdom.

[3] Keenan, Ed, Adam Czauderna, Greg Leach, Jane Cleland-Huang, Yonghee Shin, Evan Moritz, Malcom Gethers, Denys Poshyvanyk, Jonathan Maletic, Jane Huffman Hayes, Alex Dekhtyar, Daria Manukian, Shervin Hussein, Derek Hearn. "Tracelab: An experimental workbench for equipping researchers to innovate, synthesize, and comparatively evaluate traceability solutions." Proceedings of the 34th International Conference on Software Engineering. IEEE Press, 2012.

[4] Dit, Bogdan, Evan Moritz, Mario Linares-Vasquez, Denys Poshyvanyk, and Jane Cleland-Huang "Supporting and accelerating reproducible empirical research in software evolution and maintenance using TraceLab Component Library." Empirical Software Engineering (2014): 1-39.I. S. Jacobs and C. P. Bean, "Fine particles, thin films and exchange anisotropy," in Magnetism, vol. III, G. T. Rado and H. Suhl, Eds. New York: Academic, 1963, pp. 271–350.

[5] Maalej, Walid, and Hadeer Nabil. "Bug report, feature request, or simply praise? On automatically classifying app reviews." 2015 IEEE 23rd international requirements engineering conference (RE). IEEE, 2015.

[6] Hall, Mark, et al. "The WEKA data mining software: an update." ACM SIGKDD explorations newsletter 11.1 (2009): 10-18.

[7] McCallum, Andrew, and Kamal Nigam. "A comparison of event models for naive Bayes text classification." AAAI-98 workshop on learning for text categorization. Vol. 752. 1998.

[8] Bhargava, Neeraj, Ritu Bhargava, Mainsh Mathura. "Decision tree analysis on j48 algorithm for data mining." Proceedings of International Journal of Advanced Research in Computer Science and Software Engineering 3.6 (2013).

[9] Aher, Sunita B., and L. M. R. J. Lobo. "Comparative Study of Classification Algorithms." International Journal of Information Technology 5.2 (2012): 239-43.

[10] Cahoy, Ellysa. "What is Empirical Research?" PennState University Libraries. http://psu.libguides.com/emp

[11] "About TraceLab," COEST. http://www.coest.org/index.php/tracelab/about-tracelab

[12] "Apache Lucene," https://lucene.apache.org/

[13] Sayad, Saed. "ZeroR" An Introduction to Data Mining. http://www.saedsayad.com/zeror.htm

[14] Sayad, Saed. "Naïve Bayesian" An Introduction to Data Mining. http://www.saedsayad.com/naive_bayesian.htm

[15] Schneider, Jeff. "Cross Validation". https://www.cs.cmu.edu/~schneide/tut5/node42.html

[16] Hayes, Jane Huffman, Wenbin Li, and Mona Rahimi. "Weka meets TraceLab: Toward convenient classification: Machine learning for requirements engineering problems: A position paper." Artificial Intelligence for Requirements Engineering (AIRE), 2014 IEEE 1st International Workshop on. IEEE, 2014.

[17] Hayes, Jane Huffman, Wenbin Li, Tingting Yu, Xue Han, Mark Hayes, and Clinton Woodson. "Measuring Requirement Quality to Predict Testability." 2015 IEEE Second International Workshop on Artificial Intelligence for Requirements Engineering (AIRE). IEEE, 2015.